\documentclass{ws-jai}
\usepackage[flushleft]{threeparttable}
\usepackage{color}
\usepackage{hyperref}

\begin{document}
%%%%%%%%%%%%%%%%%%%%%%%
% Define new commands %
%%%%%%%%%%%%%%%%%%%%%%%
\newcommand{\mcgill}{\dagger}
\newcommand{\wvu}{\ddagger}
\newcommand{\wvucosmo}{\ast}
\newcommand{\cifar}{\S}
\newcommand{\ubc}{\star}

\newcommand{\KBcomment}[1]{#1}
\newcommand{\RScomment}[1]{{\color{red}#1}}

\catchline{}{}{}{}{} % Publisher's Area please ignore

\markboth{J.\ Mena-Parra et al.}{Quantization bias for digital correlators}

\title{Quantization bias for digital correlators}

\author{J.\ Mena-Parra$^\mcgill$, K.\ Bandura$^{\wvu,\wvucosmo}$,
M.A.\ Dobbs$^{\mcgill,\cifar}$, J.R.\ Shaw$^\ubc$, S.\ Siegel$^\mcgill$}

\address{
$^\mcgill$Department of Physics, McGill University, Montreal,
 Quebec H3A~2T8, Canada \\
$^\wvu$Lane Department of Computer Science and Electrical Engineering, 
West Virginia University, Morgantown, WV 26505 \\
$^\wvucosmo$Center for Gravitational Waves and Cosmology, 
West Virginia University, Morgantown, WV 26505 \\
$^\cifar$Canadian Institute for Advanced Research, 
Toronto, Ontario M5G~1Z8, Canada\\
$^\ubc$Department of Physics \& Astronomy, University of British Columbia, 
Vancouver, V6T 1Z1, Canada\\ 
}

\maketitle

\corres{$^\S$Send correspondence to J. Mena-Parra. E-mail: juan.menaparra@mail.mcgill.ca.}

\begin{history}
\received{(to be inserted by publisher)};
\revised{(to be inserted by publisher)};
\accepted{(to be inserted by publisher)};
\end{history}

\begin{abstract}
In radio interferometry, the quantization process introduces a bias in 
the magnitude and phase of the measured correlations which translates into 
errors in the measurement of source brightness and position in the sky,
affecting both the system calibration and image reconstruction.
In this paper we investigate the biasing effect of quantization in 
the measured correlation between complex-valued inputs with a circularly 
symmetric Gaussian probability density function (PDF), which is the typical case
for radio astronomy applications. We start by calculating the correlation 
between the input and quantization error and its effect on the quantized 
variance, first in the case of a real-valued quantizer with a zero 
mean Gaussian input and then in the case of a complex-valued quantizer with a
circularly symmetric Gaussian input. 
We demonstrate that this input-error correlation is always negative for a 
quantizer with an odd
number of levels, while for an even number of levels this correlation is 
positive in the low signal level regime. 
In both cases there is an optimal interval for the input signal level for which 
this input-error correlation is very weak and the model of additive uncorrelated 
quantization noise provides a very accurate approximation.
We determine the conditions 
under which the magnitude and phase of the measured correlation  
have negligible bias with respect to the unquantized values: we 
demonstrate that the magnitude bias is negligible only if both 
unquantized inputs are optimally quantized 
(i.e., when the uncorrelated quantization error model is valid), 
while the phase bias is negligible when 1) at least one of the  
inputs is optimally quantized, or when 2) the correlation coefficient between 
the unquantized inputs is small.
Finally, we determine the implications of these results for radio 
interferometry.
\end{abstract}


This version \LaTeX-ed \today.

%-------------------------------------------------------------------------------
%%%%% Sections %%%%%

\section{Introduction}

In radio astronomy, the digital correlator is the device that 
processes the sky signals from an interferometric array and computes
the complex-valued correlations between voltages
from pairs of inputs. These correlations provide the visibilities which
are the fundamental quantities in radio interferometry.

A digital correlator typically contains several quantization stages where the
signal amplitude is encoded with a finite set of discrete values.
In radio interferometry, this quantization process introduces a bias in the 
magnitude and phase of the measured correlations which translates into 
errors in the measurement of source brightness and position in the sky, 
affecting both the system calibration and image reconstruction. 
As we will show, this effect can be 
significant for large deviations from optimal signal levels or large correlation
coefficients, which means that
it is critical to understand \KBcomment{the bias}. This implies
understanding the statistics of the quantization error and its correlation with
the quantizer input.

In most cases the quantization error is modeled as additive
stationary white noise that has a uniform bounded distribution and is 
uncorrelated with the input. In general, this model provides a very good 
approximation when the quantization step size is small, the input signal 
traverses many quantization steps between successive 
samples and the effect of clipping (for input values outside the quantizer's
dynamic range) is small or negligible. In this case 
\citet{thompson1998} derives formulas for the fractional increase
in the variance of a white Gaussian real input signal that results from 
quantization with many levels (eight or more) and provides tables with the
optimal input signal levels that minimize this effect. Although the uncorrelated
quantization error model is still very accurate even for significant deviations
from the optimal signal level (the range depending on the number of levels), the
model breaks when the input signal level is too small or when it is too high and 
the effect of clipping is important (i.e. when the fraction of samples that lie
outside the quantizer limits is significant). More important,
it leads to the incorrect
conclusion that the magnitude and phase of the quantized correlation 
remains unbiased. We will show that this is not the case in the signal regimes 
described above and when the correlation coefficient is large.

The effect of quantization on correlators has been studied 
in the past for quantization with few levels (e.g. \citealt{1446497} for two 
levels, \citealt{kulkarni1980} for three levels, \citealt{1970AuJPh..23..521C} for 
four levels). For many levels, \citet{thompson1998} studies the loss in 
efficiency in a correlator resulting from quantization with
eight or more levels for real Gaussian inputs assuming that
the quantization error is uncorrelated with the unquantized input, while 
\citet{7771586} finds the component of the quantization noise that is 
uncorrelated with the input and calculates the loss in efficiency due to this
component. \citet{thompson2017interferometry} presents a detailed discussion
on these methods. 
Recent work from \citet{2016arXiv160804367B} 
%uses Price's theorem \cite{1057444} to
generalizes the Van Vleck quantization correction for two-level correlators
to correlators with multilevel quantization and Gaussian inputs.
Since it is not always computationally feasible to implement this correction, 
in this paper we investigate in detail the biasing effect of quantization 
on the magnitude and phase of the measured correlations and determine the
conditions under which this effect is negligible  
so the correction is not necessary. In order to do that, we  
%do not rely on Price's theorem and instead 
calculate the contribution of
each quantization level to the correlation between the 
input and the quantization error in the case of a 
single quantizer, and in the case of two quantizers with different inputs we
calculate the correlation between quantization errors for 
every pair of quantization levels\footnote{This method differs from
\citet{thompson2017interferometry} and \citet{2016arXiv160804367B} since 
it does not use Price's theorem \cite{1057444}, a very useful tool for 
estimating the expectation of nonlinear functions of jointly Gaussian random
variables. The approach used in our paper applies to generic
probability density functions, and can be used for example, to
investigate the effect of quantization in the presence of 
Radio Frequency Interference (RFI), although that analysis is beyond the
scope of this work.}. 
We then use these results to calculate the
effect of the quantization errors on the measured correlation of a real 
and complex-valued correlator.

This work is motivated by the ongoing effort to calibrate
the Canadian Hydrogen Intensity Mapping Experiment (CHIME), a hybrid 
cylindrical transit interferometer designed to measure 
the emission of 21 cm radiation from neutral hydrogen during the 
epoch when dark energy generated the transition from decelerated to accelerated 
expansion of the universe \citep{Bandura14,Newburgh14}. The most 
important challenge for CHIME comes from the calibration 
required to separate the 21 cm signal from astrophysical foregrounds that are
many orders of magnitude brighter: the proper reconstruction of 
the 21 cm power spectrum requires that all the telescope primary beams 
(direction dependent gains, fairly stable in time) are known to $\sim 0.1\%$ 
and the receiver gains (direction independent, vary with time) to $\sim 1\%$ on 
short time scales \cite{2015PhRvD..91h3514S}. The bias in the 
correlations due to quantization will show as an amplitude dependent 
(and direction independent) gain term that must be addressed before beam and 
receiver gain calibration.

The CHIME correlator is based on an FX design, where the F-engine digitizes
(samples at 800 MSPS and quantizes to 8 bits) and channelizes (i.e., divide the 
400 MHz input bandwidth into thousands of frequency channels) the analog signals
from the 2048 receivers of the interferometer
\citep[see][for details of the CHIME F-engine]{2016JAI.....541005B}. 
The complex-valued data from each frequency channel is then quantized 
to 4 bits (4 bits real + 4 bits imaginary)
before being reorganized by a corner-turn system 
\citep[see][for details of the corner-turn network]{2016JAI.....541004B} and fed 
into the X-engine that computes the full $N^2$ correlation matrix 
\citep[see][for details of the CHIME X-engine]{Denman:2015ec,Recnik:2015ev}.
The CHIME correlator is currently the largest radio correlator that has been 
built as measured in number of inputs squared times bandwidth.
Although the results of this paper are general and apply to any digital 
correlator, we will focus our analysis and simulations mainly on the 
4-bit real + 4-bit imaginary complex-valued quantization at the channelization 
stage of the CHIME correlator. We will refer to the CHIME case
to explain the implications of our results for radio interferometry.
We are particularly interested
in the effects of quantization in the \KBcomment{high signal level
and high correlation regimes which are} relevant
for CHIME when the antenna temperature and thus the correlator input signal can
increase significantly relative to the optimal level (typically determined at 
night hours or when observing a relatively quiet region of the sky), for example
during bright point source transits (e.g. the sun) and point-source calibration, 
and during complex receiver gain calibration where a broadband and relatively
bright (with a signal-to-system-noise ratio that can exceed -10 dB) 
noise-like signal is injected across the array \citep{Newburgh14}.

The layout of this document is as follows: In Section \ref{sec:real_quantizer}
we investigate the quantizer behavior and the correlation between the 
unquantized input and the quantization error for the nominal case of 
a real-valued (independent and identically distributed, IID) Gaussian input. 
In Section \ref{sec:complex_quantizer} we extend the results 
to the case of a complex-valued quantizer with a complex 
circularly symmetric Gaussian input. In Section \ref{sec:real_correlator}
we investigate the effect of quantization on the measured correlation  
between two real-valued inputs that have a joint Gaussian distribution. In 
Section \ref{sec:complex_correlator} we extend to the case of a complex-valued 
correlator and establish the conditions under which the magnitude and phase of
the measured correlation have negligible bias. In Section 
\ref{sec:interferometry} we determine the implications of these results for radio 
interferometry.

%###############################################################################
\section{Real-valued quantizer}
\label{sec:real_quantizer}

We will assume a quantizer with uniformly spaced levels and an odd 
symmetric transfer function (same number of levels above and below zero). 
This means that if the number of levels $N$ is odd then the quantizer has a 
has a level at zero (mid-tread) and if $N$ is even it has a threshold at zero 
(mid-riser). 
\KBcomment{We do not consider non-uniform quantization steps for optimization.
The CHIME case, which we assume as an example, 
corresponds to $N=15$ (levels at -7, -6, ..., 6, 7) for the complex 
channelization stage.} 
In general, the quantizer levels are (in units of the 
quantization step $\Delta$)

\begin{equation}
\label{eq:q_levels}
    k_i = -\frac{N+1}{2}+i, \text{ for } i=1, ..., N
\end{equation}

\noindent and the decision thresholds are

\begin{equation}
\label{eq:q_thresholds}
    y_0 = -\infty, \hspace{0.5in} y_N= \infty, \hspace{0.5in}
    y_i = k_i+\frac{1}{2}=-\frac{N}{2}+i, \text{ for } i=1, ..., N-1.
\end{equation}

Let $v$ be the (real-valued) input of the quantizer. For the $i$-th quantization
level, the correlation between the input and the quantization error is 
\citep{31003}

\begin{equation}
\label{eq:r_ve_i}
    \langle ve \rangle_i = \int_{y_{i-1}}^{y_i} (k_i-v)v f(v) dv
\end{equation}

\noindent where $v$ is in quantization step units and has probability density 
function (PDF) $f(v)$. Each input sample can fall in only one 
quantization slot so events that take place in the various slots are mutually 
exclusive. This means that we can write 
the correlation between the input $v$ and the quantization error $e=\hat{v}-v$
as ($\hat{v}$ is the quantizer output)

\begin{equation}
\label{eq:r_ve}
    \langle ve \rangle = \sum_{i=1}^N \int_{y_{i-1}}^{y_i} (k_i-v)v f(v) dv.
\end{equation}

Similarly, the quantization error variance $\sigma_e^2 = \langle e^2 \rangle$ 
can be written as

\begin{equation}
\label{eq:var_e_general}
    \sigma_e^2 = \sum_{i=1}^N \int_{y_{i-1}}^{y_i} (k_i-v)^2 f(v) dv.
\end{equation}

As Equation \ref{eq:r_ve} shows, the calculation of $\langle ve \rangle$ 
depends on the input PDF. If $v$ is an IID Gaussian process with zero mean, then
Equation \ref{eq:r_ve} can be written in a more concrete form

\begin{equation}
\label{eq:r_ve_1}
\begin{split}
    \langle ve \rangle 
    & = \sum_{i=1}^N \int_{y_{i-1}}^{y_i} (k_i-v)v \mathcal{N}(v|\sigma^2) dv
      = \sigma^2 \left[-1 + \sum_{i=1}^{N-1} 
            \mathcal{N}\left(\left.-\frac{N}{2}+i\right|\sigma^2\right)\right]\\
    & = \begin{cases} \displaystyle
        \sigma^2 \left[-1 + 2\sum_{i=0}^{\frac{N-3}{2}} 
            \mathcal{N}\left(\left.\frac{1}{2}+i\right|\sigma^2\right) \right]
            & \text{if $N$ odd}\\ \displaystyle
        \sigma^2 \left[-1 + \frac{1}{\sqrt{2\pi\sigma^2}} + 2\sum_{i=0}^{\frac{N-4}{2}} 
            \mathcal{N}\left(\left.1+i\right|\sigma^2\right) \right]
            & \text{if $N$ even}  
        \end{cases}          
\end{split}
\end{equation}

\noindent where 
$\mathcal{N}(v|\sigma^2)=\left(2\pi\sigma^2\right)^{-1/2}e^{-v^2/(2\sigma^2)}$
is the Gaussian PDF, 
$\sigma$ is in units of the quantization step $\Delta$,
and it is clear that the summation term is 
zero for $N=2$. Appendix \hyperref[app:r_ve]{A} provides a derivation 
for Equation \ref{eq:r_ve_1}.

It is also clear from the symmetry of the
quantizer and the input PDF that both $e$ and $\hat{v}$ have zero mean. 
Using the same procedure we find the variance of the
quantization error $\sigma_e^2 = \langle e^2 \rangle$ as
(see Appendix \hyperref[app:r_ve]{A} for details)

\begin{equation}
\label{eq:var_e}
\begin{split}
    \sigma_e^2
    & = \sum_{i=1}^N \int_{y_{i-1}}^{y_i} (k_i-v)^2 \mathcal{N}(v|\sigma^2)dv \
      = -2\langle ve\rangle -\sigma^2 + \left(\frac{N-1}{2}\right)^2
        - \sum_{i=1}^{N-1} \left(-\frac{N}{2}+i\right)          
       \text{erf}\left(\frac{-N/2+i}{\sqrt{2\sigma^2}}\right).
\end{split}
\end{equation}

\noindent where erf($v$) is the error function. 
The quantized output variance $\hat{\sigma}^2$ follows from Equations
\ref{eq:r_ve_1} and \ref{eq:var_e}

\begin{equation}
\label{eq:var_vo}
\begin{split}
    \hat{\sigma}^2 
    & = \langle (v+e)^2\rangle
    = \sigma_e^2+\sigma^2+2\langle ve\rangle 
    = \left(\frac{N-1}{2}\right)^2
        - \sum_{i=1}^{N-1} \left(-\frac{N}{2}+i\right)          
       \text{erf}\left(\frac{-N/2+i}{\sqrt{2\sigma^2}}\right)\\
    & = \begin{cases} \displaystyle
        \left(\frac{N-1}{2}\right)^2 - 2\sum_{i=0}^{\frac{N-3}{2}} 
            \left(\frac{1}{2}+i\right)          
            \text{erf}\left(\frac{1/2+i}{\sqrt{2\sigma^2}}\right)
            & \text{if $N$ odd}\\ \displaystyle
        \left(\frac{N-1}{2}\right)^2 - 2\sum_{i=0}^{\frac{N-4}{2}} 
            \left(1+i\right)          
            \text{erf}\left(\frac{1+i}{\sqrt{2\sigma^2}}\right)
            & \text{if $N$ even.}  
        \end{cases}          
\end{split}
\end{equation}

\begin{figure}[htbp]
    \centering
    \includegraphics[width=0.9\textwidth]{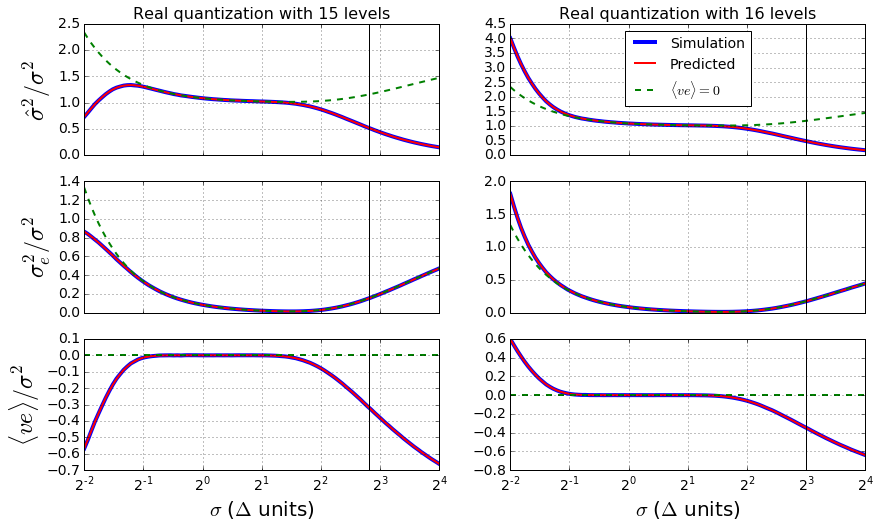}
    \caption{
Behavior of a quantizer with $N=15$ levels 
(left column) and $N=16$ levels (right column) and a real-valued Gaussian input. 
From top to bottom row, the plots show the 
variance  of the quantized output, $\hat{\sigma}^2$, the
quantization error, $\sigma_e^2$, and the correlation  between the input and
quantization error, $\langle ve \rangle$,
as function of the unquantized 
standard deviation $\sigma$. All the values are normalized with respect to 
$\sigma^2$. For each plot, the red line
corresponds to Equations~\ref{eq:r_ve_1} - \ref{eq:var_vo}, 
\KBcomment{the thick blue line shows}
the results from simulations, and the green dashed line corresponds to
the uncorrelated quantization noise model that assumes $\langle ve \rangle=0$.
Note that Equations \ref{eq:r_ve_1} - \ref{eq:var_vo} predict accurately the 
results from simulations. When the input $\sigma$ uses optimally the quantizer's
dynamic range the quantization error is very weakly correlated with the input.
In this case the uncorrelated quantization noise model provides 
a very good approximation, introducing only a small bias error.
    }
    \label{fig:real_quantization_15_16_levels}
\end{figure}

%Equation \ref{eq:var_vo} is equivalent to Equation 34 in 
%\cite{2016arXiv160804367B} for $N$ odd.
Results from simulations where we verify Equations~
\ref{eq:r_ve_1} - \ref{eq:var_vo} for the case of a real quantizer with $N=15$
levels (left column) and $N=16$ levels (right column)
and a real Gaussian input are shown in Figure 
\ref{fig:real_quantization_15_16_levels}. From top to bottom row, the plots show 
the variance of the quantized output, $\hat{\sigma}^2$, the
quantization error, $\sigma_e^2$, and the correlation between the input and
quantization error $\langle ve \rangle$
as function of the unquantized 
standard deviation $\sigma$. All the values are normalized with respect to 
$\sigma^2$. For easier visualization of the results, especially
in the low and high signal level regimes, 
the x-axis is in logarithmic scale (base 2, so the exponents can be interpeted 
as bits rms). 
For each plot, the red line
corresponds to Equations~\ref{eq:r_ve_1} - \ref{eq:var_vo} and the blue line 
\KBcomment{(made thicker so it can be distinguished from the red line)}
corresponds to the results from simulations where, for each value of $\sigma$, 
$10^6$ samples of a Gaussian input are quantized with $N$ levels and then the 
statistics of the input, output and quantization error are calculated.
As reference, we also include the green dashed line that 
shows to the expected behavior from the uncorrelated quantization noise model 
that assumes $\langle ve \rangle=0$ 
\citep[see][for a detailed discussion]{thompson2017interferometry}. 
The black solid vertical line corresponds to the highest level
of the quantizer (7 for $N=15$ and 8 for $N=16$) above which clipping occurs.

The first thing to note from Figure \ref{fig:real_quantization_15_16_levels} is  
that Equations \ref{eq:r_ve_1} - \ref{eq:var_vo} predict accurately the results 
from simulations regarding $\hat{\sigma}^2,~\sigma_e^2$, and 
$\langle ve \rangle$.
Also that the uncorrelated quantization noise model provides 
an excellent approximation in the interval where 
$\langle ve \rangle \rightarrow 0$.
For $N=15$, the value of $\sigma$ that minimizes the magnitude of the 
input-error correlation coefficient, 
$\rho_{ve}=\langle ve \rangle/(\sigma\sigma_e)$, is $\sigma \approx 2^{0.14} \Delta$. At
this point $|\rho_{ve}| \approx 5.5 \times 10^{-10}$. For $N=16$ we have
$\rho_{ve}=0$ for $\sigma \approx 2^{0.2} \Delta$. In both cases the minimum of
$|\rho_{ve}|$ is broad so there is effectively a $\sigma$-interval, which 
we denote the interval of optimal quantization, for which the correlation 
between the input and quantization error is very weak and the 
uncorrelated quantization error model provides a very accurate approximation
(the error in the calculated quantization parameters is negligible).
The length of this interval depends on $N$ and on the tolerance
required by each specific application. For example, if we require that
$|\rho_{ve}|\lesssim 10^{-3}$ for $N=15$, then the interval of optimal 
quantization is, approximately, $[2^{-0.6}, 2^{0.9}]$. Within this interval the values of 
$\hat{\sigma}^2$ and $\sigma_e^2$ from the uncorrelated quantization error model
agree with the values from Equations \ref{eq:r_ve_1} - \ref{eq:var_vo} at the
$\sim 0.07\%$ level. The performance of the $N=16$ quantizer within this interval is
similar\footnote{The CHIME correlator also has a quantizer with $N=255$
levels at the digitization stage. For this quantizer the interval of optimal
quantization is much broader, spanning several bits, and the correlation 
between the input and quantization error over this interval is even weaker
$\left(|\rho_{ve}|\ll 10^{-14}\right)$. The effects of this correlation are negligible
compared to the $N=16$ complex-valued quantizer at the channelization stage.}. 

Also note that, even in the high-$\sigma$ regime, where the quantization error 
resulting from clipping dominates and is correlated with the input, 
the uncorrelated quantization noise model also predicts with high accuracy the 
contribution of this overload error to $\sigma_e^2$
as the middle plot shows. 
However, it cannot track the quantized standard deviation (top plot)
since in this regime $\langle ve \rangle<0$ which eventually makes 
$\hat{\sigma}^2/\sigma^2<1$ for large inputs. 
In the low-$\sigma$ regime, when $\sigma \lesssim 1/2$, the 
uncorrelated quantization noise model deviates from Equations 
\ref{eq:r_ve_1} - \ref{eq:var_vo} for two reasons: first, it is no longer 
true that the quantization error is uniformly distributed in the 
interval $[-\Delta/2, \Delta/2]$, and second, the behavior is now closer to 
that of a 3-bit ($N$ odd) or 2-bit ($N$ even) quantizer, so the quantization 
error is again correlated with the input. As $N$ increases, both the interval of
optimal quantization and the accuracy of the uncorrelated quantization error 
model increase.

Finally, note that $\langle ve \rangle$ is negative 
(it approaches zero assymptotically) for $N=15$ while it 
becomes positive in the low signal level regime for $N=16$. 
Since the sum 
$S_o = \sum_{i=0}^{(N-3)/2} \mathcal{N}\left(\left. 1/2+i\right|\sigma^2\right)$
in Equation \ref{eq:r_ve_1} is positive and
bounded above by 1/2 ($S_o < 1/2$, a proof is provided in
Appendix \ref{app:sign_r_ve}) 
then $\langle ve \rangle$ is always negative for $N$ odd. 
Furthermore, $\langle ve \rangle \in (-\sigma^2, 0)$ in this case.
On the other hand, for $N$ even, the sum
$S_e = \sum_{i=0}^{(N-4)/2} \mathcal{N}\left(\left. 1+i\right|\sigma^2\right)$
is also positive and bounded above by 1/2, 
but the term $1/(\sqrt{2\pi\sigma^2})$ becomes 
arbitrarily large as $\sigma$ decreases. Thus, $\langle ve \rangle$ is always 
positive and unbounded for $N$ even in the low-$\sigma$ regime.

%###############################################################################
\section{Complex-valued quantizer}
\label{sec:complex_quantizer}

In the CHIME correlator, the (real-valued) analog signal of each input is first 
digitized and then passed through the F-engine that implements a Polyphase
Filter Bank (PFB) which splits the 400 MHz-wide input into 1024 frequency bins, 
each 390 kHz wide. The output of each frequency bin is a complex-valued signal  
and its real and imaginary parts are separately quantized with 15 levels before 
the data is re-arranged and sent to the X-engine for cross-multiplication and
integration. In this section we extend the results of Section 
\ref{sec:real_quantizer} to the case of an $N$-level complex-valued 
quantizer, where the real and imaginary parts of the input are separately 
quantized with $N$ levels. In this case we assume that the input $v=v_r+jv_i$ 
is a complex and circularly-symmetric Gaussian process such that 
$\langle v_r v_i\rangle=0$ and 
$\langle v_r^2\rangle=\langle v_i^2\rangle=\langle |v|^2\rangle/2$ where
$\langle |v|^2\rangle=\sigma^2$ is the the unquantized standard deviation. 
As in Section \ref{sec:real_quantizer} we are interested in the standard
deviation of the quantization error, $e=e_r+je_i$, and its correlation with the 
input. In this case we have

\begin{equation}
\label{eq:r_ve_complex}
\begin{split}
    \langle v e^* \rangle
    &= \langle (v_r+jv_i) (e_r+je_i)^* \rangle \\
    &= \langle v_r e_r \rangle+ \langle v_i e_i \rangle + 
    j \left ( -\langle v_r e_i \rangle +  \langle v_i e_r \rangle \right ).
\end{split}
\end{equation}

The circular symmetry of $v$ (its real and imaginary part are uncorrelated and 
have identical statistics) implies that 
 $\langle v_r e_r \rangle=\langle v_i e_i \rangle$. As for  
 $\langle v_r e_i \rangle$, note that, for the $m$-th imaginary quantization 
 level we have 

\begin{equation}
\label{eq:r_vr_ei_m}
\begin{split}
    \langle v_re_i \rangle_m 
    & = \int_{y_{m-1}}^{y_m} \int_{-\infty}^\infty 
        (k_m-v_i)v_r f(v_r, v_i) dv_r dv_i\\
    & = \int_{y_{m-1}}^{y_m} (k_m-v_i) 
        \mathcal{N}\left(v_i\left|\frac{\sigma^2}{2}\right.\right)dv_i
        \int_{-\infty}^\infty 
        v_r\mathcal{N}\left(v_r\left|\frac{\sigma^2}{2}\right.\right) 
        dv_r \\
    & = 0.
\end{split}
\end{equation}

Thus $\langle v_r e_i \rangle=0$ and, for the same reason, 
$\langle v_i e_r \rangle=0$. This means that 
 $\langle v e^* \rangle$ is real and, from Equation \ref{eq:r_ve_1}

 \begin{equation}
\label{eq:r_ve_complex_1}
    \langle v e^* \rangle 
    = 2\langle v_r e_r \rangle 
    = \sigma^2 \left[-1 + \sum_{i=1}^{N-1} 
   \mathcal{N}\left(\left.-\frac{N}{2}+i\right|\frac{\sigma^2}{2}\right)\right].
\end{equation}

From the circular symmetry of $v$ it also follows that 
$\sigma_e^2 = 2\langle e_r^2 \rangle$ and 
$\hat{\sigma}^2 = 2\langle \hat{v}_r^2 \rangle$, so similar expressions
for $\sigma_e^2/\sigma^2$ and $\hat{\sigma}^2/\sigma^2$ in the
complex case are obtained from  
Equations~\ref{eq:r_ve_1} - \ref{eq:var_vo}
by changing $\sigma^2 \rightarrow \sigma^2/2$.

\begin{figure}[htbp]
    \centering
    \includegraphics[width=\textwidth]{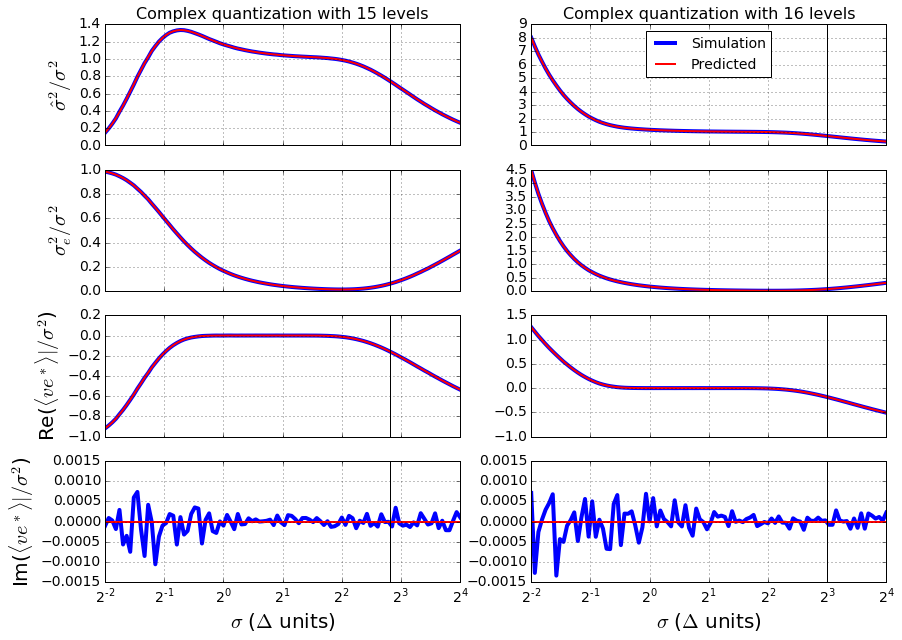}
    \caption{
Behavior of a complex-valued quantizer with $N=15$ levels 
(left column) and $N=16$ levels (right column) and a 
circularly-symmetric Gaussian input. 
From top to bottom row, the plots show the 
normalized variance of the quantized output ($\hat{\sigma}^2/\sigma^2$), 
the quantization error ($\sigma_e^2/\sigma^2$), and the magnitude and phase
(in degrees) of the normalized 
 correlation between the input and quantization error 
($\langle ve^* \rangle/\sigma^2$).
For each plot, the red line is our prediction 
and the blue line is the result from simulations. There is again excellent 
agreement between these. Note that $\langle ve^* \rangle$ is always real
\KBcomment{(in the simulation the imaginary part is consistent with zero at  
the $\sim$~0.15\% level)}, and
it is negative ($180^\circ$ phase) for $N$ odd, while it becomes positive 
($0^\circ$ phase) in the low $\sigma$ regime for $N$ even.
    }
    \label{fig:complex_quantization_15_16_levels}
\end{figure}

Results from simulations and comparison to our prediction for the complex-valued
quantizer with $N=15$ levels (left column) and $N=16$ levels (right column) are 
shown in Figure \ref{fig:complex_quantization_15_16_levels}. 
From top to bottom row, the plots show the 
normalized variance of the quantized output ($\hat{\sigma}^2/\sigma^2$), 
the quantization error ($\sigma_e^2/\sigma^2$), and the magnitude and phase
(in degrees) of the normalized 
 correlation between the input and quantization error 
($\langle ve^* \rangle/\sigma^2$).
For each plot, the red line is our prediction 
and the blue line corresponds to the results from simulations where, for each 
value of $\sigma$, $10^6$ samples of a complex and 
circularly-symmetric Gaussian input are quantized with $N$ levels (real and
imaginary parts quantized separately) and then the 
statistics of the input, output and quantization error are calculated.

There is again excellent agreement between the simulations and the 
predictions. The correlation between the input and quantization error, 
$\langle ve^* \rangle$, is 
always real \KBcomment{(in the simulation the imaginary part is consistent with
zero at the $\sim$~0.15\% level)}. 
Furthermore, it is always negative ($180^\circ$ phase) for $N$ odd, 
while it becomes positive ($0^\circ$ phase) in the low $\sigma$ regime for $N$ 
even. In this case the optimal quantization interval corresponding
to $|\rho_{ve}|\lesssim 10^{-3}$ is approximately $[2^{-0.1}, 2^{1.4}]$
(the interval shifts by $\sqrt{2}$ with respect to the real-valued case).

%###############################################################################
\section{Real-valued correlator}
\label{sec:real_correlator}

The correlation between two real-valued quantized inputs 
$\hat{v}_1$ and $\hat{v}_2$, is 

\begin{equation}
\label{eq:r}
    \hat{r}_{12}=\langle \hat{v}_1\hat{v}_2 \rangle.
\end{equation}

The output of a real-valued digital correlator after integrating $N_s$ samples 
is

\begin{equation}
\label{eq:r_Ns}
    \hat{r}_{12,N_s} = \frac{1}{N_s}\sum_{n=1}^{N_s} \hat{v}_1[n]\hat{v}_2[n].
\end{equation}

Since the quantized sample vector $(\hat{v}_1[n], \hat{v}_2[n])$ comes from
the IID joint Gaussian process $(v_1, v_2)$, then 
$\langle \hat{r}_{12,N_s} \rangle = \hat{r}_{12}$ so
the measured correlation $\hat{r}_{12,N_s}$ is an
unbiased estimator of $\hat{r}_{12}$. 
Henceforth we will refer to $\hat{r}_{12}$ as the output of the
digital correlator.

Note that we already investigated the behavior of 
$\hat{r}_{11}=\hat{\sigma}_1^2$ and $\hat{r}_{22}=\hat{\sigma}_2^2$ in Section 
\ref{sec:real_quantizer} (the result in this case is the same because the 
marginal PDFs of $v_1$ and $v_2$ 
are independent of the correlation between inputs). Now we are 
interested in $\hat{r}_{12}$ and its relation to 
$r_{12}=\langle v_1 v_2 \rangle$ which is the 
correlation between the unquantized inputs $v_1$ and $v_2$ and what we 
ultimately want to measure.
We can write $\hat{r}_{12}$ as

\begin{equation}
\label{eq:r_12}
\begin{split}
    \hat{r}_{12}
    & = \langle (v_1+e_1)(v_2+e_2) \rangle \\
    & = r_{12} + \langle v_1e_2 \rangle + \langle e_1v_2 \rangle + 
        \langle e_1e_2 \rangle
\end{split}
\end{equation}

\noindent where

\begin{equation}
\label{eq:r_v1_e2}
\begin{split}
    \langle v_1e_2 \rangle 
    & = \sum_{i=1}^{N}\int_{y_{i-1}}^{y_i} \int_{-\infty}^\infty 
        (k_i-v_2)v_1 f(v_1, v_2) dv_1 dv_2\\
\end{split}
\end{equation}
 
\noindent and

\begin{equation}
\label{eq:r_e1_e2}
\begin{split}
    \langle e_1e_2 \rangle 
    & = \sum_{j=1}^{N}\sum_{i=1}^{N}\int_{y_{j-1}}^{y_j} \int_{y_{i-1}}^{y_i}
        (k_i-v_1)(k_j-v_2) f(v_1, v_2) dv_1 dv_2.
\end{split}
\end{equation}

$\langle e_1v_2 \rangle $ is defined as in Equation \ref{eq:r_v1_e2}. If the
samples from $v_1$ and $v_2$ come from a zero-mean joint Gaussian PDF

\begin{equation}
\label{eq:joint_gaussian_pdf}
    \mathcal{N}\left(v_1, v_2\left|\sigma_1^2, \sigma_2^2, \rho \right. \right)= 
    \frac{1}{2\pi\sigma_1\sigma_2\sqrt{1-\rho^2}}
    e^{-\frac{1}{2(1-\rho^2)}\left[\frac{v_1^2}{\sigma_1^2} + 
                                   \frac{v_2^2}{\sigma_2^2} -
                                   \frac{2\rho v_1 v_2}{\sigma_1\sigma_2}
                             \right]}
\end{equation}

\noindent where $\rho= \langle v_1v_2 \rangle/(\sigma_1\sigma_2)$, then Equation 
\ref{eq:r_v1_e2} can be simplified

\begin{equation}
\label{eq:r_v1_e2_1}
\begin{split}
    \langle v_1e_2 \rangle 
    & = \sum_{i=1}^{N}\int_{y_{i-1}}^{y_i} \int_{-\infty}^\infty 
        (k_i-v_2)v_1 
        \mathcal{N}\left(v_1, v_2\left|\sigma_1^2, \sigma_2^2, \rho \right. 
        \right) dv_1 dv_2\\
    & = \rho\frac{\sigma_1}{\sigma_2}\sum_{i=1}^{N}\int_{y_{i-1}}^{y_i} 
        (k_i-v_2)v_2 
        \mathcal{N}\left(v_2|\sigma_2^2\right) dv_2\\
    & = \rho\frac{\sigma_1}{\sigma_2} \langle v_2e_2 \rangle \\
    & = r_{12}\frac{\langle v_2e_2 \rangle }{\sigma_2^2}.
\end{split}
\end{equation}

Similarly $\langle e_1v_2 \rangle = r_{12}\langle v_1e_1 \rangle/\sigma_1^2$. Note 
that with this result we can find both $\langle v_1e_2 \rangle$ and 
$\langle e_1v_2 \rangle$, which are correlations between mixed input-error 
terms, using Equation \ref{eq:r_ve_1} for the correlation between an input and 
its respective quantization error.

As for $\langle e_1e_2 \rangle$ in Equation \ref{eq:r_e1_e2}, it can be 
simplified in the case
when $\rho$ is small, since in this regime we have

\begin{equation}
\label{eq:joint_gaussian_pdf_small_rho}
    \left.\mathcal{N}\left(v_1, v_2\left|\sigma_1^2, \sigma_2^2, \rho \right. 
    \right) \right |_{\rho \ll 1} \approx 
    \mathcal{N}(v_1|\sigma_1^2) \mathcal{N}(v_2|\sigma_2^2)
    \left(1+\frac{\rho v_1 v_2}{\sigma_1\sigma_2} \right)
\end{equation}

\noindent so

\begin{equation}
\label{eq:r_e1_e2_small_rho}
\begin{split}
    \left.\langle e_1e_2 \rangle \right|_{\rho \ll 1}  \approx &
    \sum_{i=1}^N \int_{y_{i-1}}^{y_i}(k_i-v_1)\mathcal{N}(v_1|\sigma_1^2)dv_1
    \sum_{j=1}^N \int_{y_{j-1}}^{y_j}(k_j-v_2)\mathcal{N}(v_2|\sigma_2^2)dv_2\\
      & + \frac{\rho}{\sigma_1\sigma_2}
    \sum_{i=1}^N \int_{y_{i-1}}^{y_i}(k_i-v_1)v_1\mathcal{N}(v_1|\sigma_1^2)dv_1
    \sum_{j=1}^N \int_{y_{j-1}}^{y_j}(k_j-v_2)v_2\mathcal{N}(v_2|\sigma_2^2)dv_2
      \\
    = & \langle e_1 \rangle\langle e_2 \rangle + 
    \frac{\rho}{\sigma_1\sigma_2}\langle v_1e_1 \rangle\langle v_2e_2 \rangle\\
    = & r_{12}\frac{\langle v_1e_1 \rangle }{\sigma_1^2}
    \frac{\langle v_2e_2 \rangle }{\sigma_2^2}.
\end{split}
\end{equation}

Equation \ref{eq:r_e1_e2_small_rho} will be useful when we analyze the phase
behavior of the complex-valued correlator. 

From Equations \ref{eq:r_12} and \ref{eq:r_v1_e2_1} we can write

\begin{equation}
\label{eq:r_12_a}
\begin{split}
    \hat{r}_{12}
    & = r_{12} \left(1 + \frac{\langle v_1e_1 \rangle }{\sigma_1^2} + 
                     \frac{\langle v_2e_2 \rangle }{\sigma_2^2} \right)
        + \langle e_1e_2 \rangle
\end{split}
\end{equation}

\noindent and, using Equation \ref{eq:r_e1_e2_small_rho}

\begin{equation}
\label{eq:r_12_small_rho}
\begin{split}
    \left. \hat{r}_{12} \right|_{\rho \ll 1}  \approx &
    ~r_{12} \left(1 + \frac{\langle v_1e_1 \rangle }{\sigma_1^2} + 
                      \frac{\langle v_2e_2 \rangle }{\sigma_2^2} +
                      \frac{\langle v_1e_1 \rangle }{\sigma_1^2}
                      \frac{\langle v_2e_2 \rangle }{\sigma_2^2} \right)
    = r_{12}\left( 1 + \frac{\langle v_1e_1 \rangle }{\sigma_1^2}\right)
    \left( 1 + \frac{\langle v_2e_2 \rangle }{\sigma_2^2}\right).
\end{split}
\end{equation}

\begin{figure}[htbp]
    \centering
    \includegraphics[width=\textwidth]{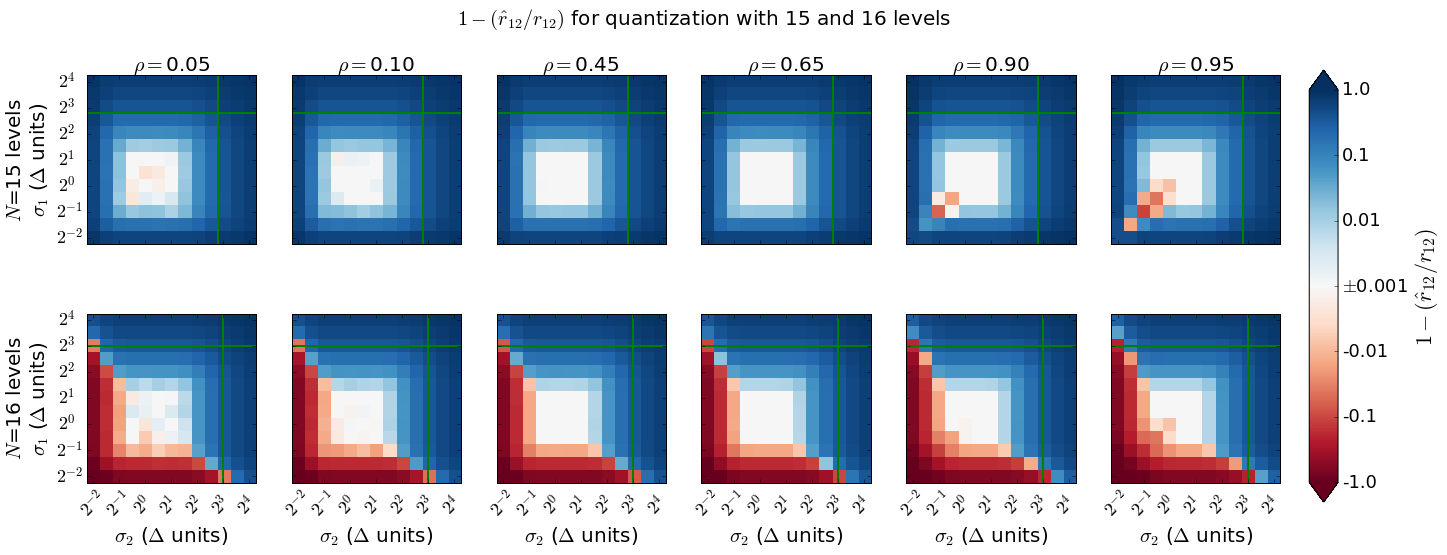}
    \caption{
Results from simulations of $r = \hat{r}_{12}/r_{12}$ 
as function of 
$\sigma_1$ and $\sigma_2$ for different values of $\rho$ 
for a real
correlator with $N=15$ levels (top row) and $N=16$ levels (bottom row).
The axes for each plot are the unquantized input signal levels and the green 
solid lines correspond to the highest level of the quantizer 
above which clipping occurs. 
The bias in $\hat{r}_{12}$ for moderate values of $\rho$ 
($|\rho|\lesssim 0.85$) is below $\sim 0.1\%$ approximately 
within the inner white square enclosed by the
region $\sigma_1\times\sigma_2 \approx [2^{-0.6}, 2^{0.9}] \times 
[2^{-0.6}, 2^{0.9}]$.
For $|\rho|\gtrsim 0.85$ the bias can increase up to 
$\sim 4\%$.
    }
    \label{fig:real_corr_15_16_levels}
\end{figure}

\begin{figure}[htbp]
    \centering
    \includegraphics[width=\textwidth]{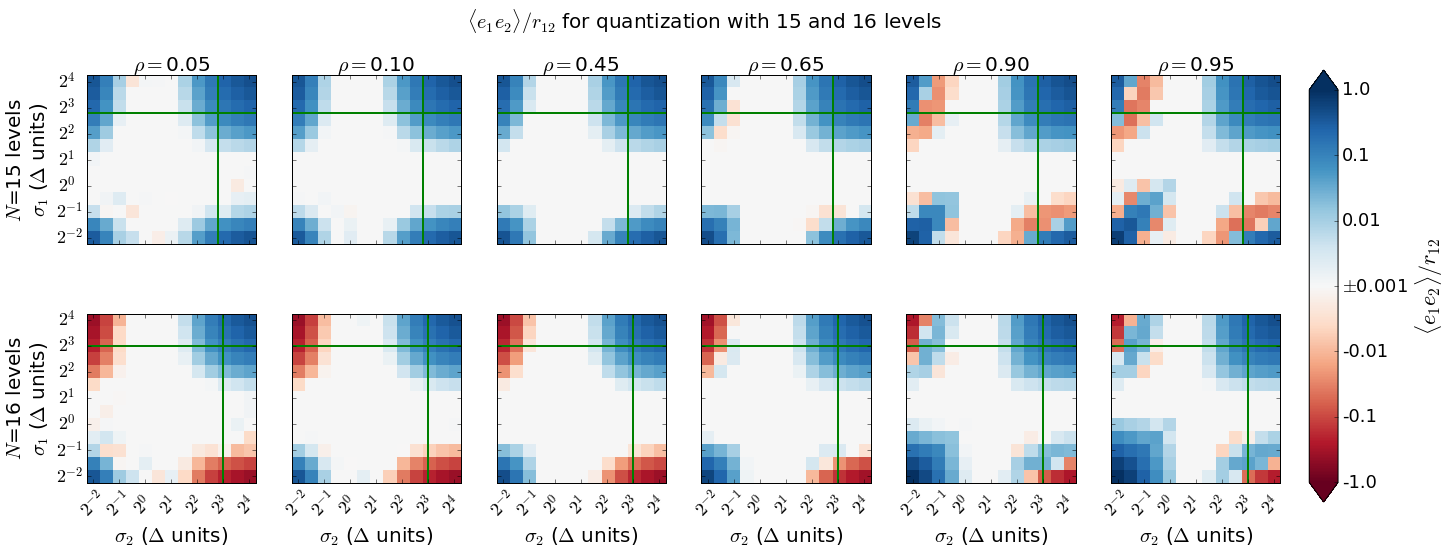}
    \caption{
Correlation between the quantization errors of the two inputs, 
$\langle e_1e_2 \rangle$ (normalized by $r_{12}$), as function of 
$\sigma_1, ~\sigma_2$ and $\rho$, from simulations. Note that
$e_1$ and $e_2$ are very weakly correlated as long as at least one of the
two inputs is optimally quantized.
    }
    \label{fig:real_corr_e1e2_15_16_levels}
\end{figure}

The behavior from simulations of the normalized and quantized input correlation 
$r = \hat{r}_{12}/r_{12}$ and the contribution of the correlation between the 
quantization errors of the two inputs, $\langle e_1e_2 \rangle$ 
(also normalized by $r_{12}$) are shown in Figures 
\ref{fig:real_corr_15_16_levels} and \ref{fig:real_corr_e1e2_15_16_levels}
respectively. For each value of $\sigma_1, ~\sigma_2$ and $\rho$, $10^7$ 
sample vectors $(v_1[n], v_2[n])$ from the joint Gaussian distribution in 
Equation \ref{eq:joint_gaussian_pdf} are quantized with $N$ levels and then both 
$\hat{r}_{12}$ and $\langle e_1e_2 \rangle$ are calculated and normalized by 
the (measured) unquantized input correlation $r_{12}$. The axes for 
each plot are the unquantized input signal levels and the green solid lines 
correspond to the highest level of the quantizer (7 for $N=15$ and 8 for $N=16$) 
above which clipping occurs.

With $10^7$ samples, the values in each pixel of Figure
\ref{fig:real_corr_15_16_levels} agree with the values from 
Equation~\ref{eq:r_12_a} with unbiased error fluctuations below $\sim$~1\%.
The worst case corresponds to low values of $\sigma$ 
and $\rho$ where $r_{12}$ is very small.
These results confirm that Equation~\ref{eq:r_12_a} accurately
reproduces the relation between $\hat{r}_{12}$ and $r_{12}$ for the 
real-valued correlator.

For moderate values of $\rho$ ($|\rho|\lesssim 0.85$)
the bias in $\hat{r}_{12}$ (Figure \ref{fig:real_corr_15_16_levels}) 
is below $\sim 0.1\%$ (values from Equation~\ref{eq:r_12_a}) approximately 
within the inner white square enclosed by the
region $\sigma_1\times\sigma_2 \approx [2^{-0.6}, 2^{0.9}] \times 
[2^{-0.6}, 2^{0.9}]$, corresponding to the region where both inputs are 
optimally quantized (see Section \ref{sec:real_quantizer}).
For $|\rho|\gtrsim 0.85$ the bias within this region can increase up to $\sim 4\%$.

The most important feature from Figure \ref{fig:real_corr_e1e2_15_16_levels}
is that $e_1$ and $e_2$ are weakly correlated 
as long as at least one of the two
inputs is approximately uncorrelated with its respective quantization 
error (either $\langle v_1e_1 \rangle$ or $\langle v_2e_2 \rangle$ is 
negligible). Another way to
say this is that $e_1$ and $e_2$ are weakly correlated as long as at 
least one of the two inputs is optimally quantized, i.e.,
when the model of additive uncorrelated quantization noise is 
(approximately) valid. Note that this is what 
one would intuitively assume using the nominal model of additive uncorrelated 
quantization error.
We will use this result when we analyze the phase of the measured correlation 
in a complex correlator.

%###############################################################################
\section{Complex-valued correlator}
\label{sec:complex_correlator}

Now we extend the results from Section \ref{sec:real_correlator} to the case 
when the correlator inputs
are complex-valued, such as for the complex channelization stage of 
the CHIME correlator, where the digitized inputs are channelized using a 
PFB that splits the 
400 MHz-wide input into 1024 narrow frequency bins. The complex-valued
output of each frequency bin is quantized with $N=15$ levels for both the
real and imaginary parts. Finally, the quantized signals are sent to
the correlator that measures complex-valued correlation between quantized 
inputs, $\hat{r}_{12} = \langle \hat{v}_1\hat{v}_2^* \rangle$. We are ultimately
interested in $r_{12} = \langle v_1v_2^* \rangle$ so we need to find a relation
between these. 

As in Section \ref{sec:complex_quantizer}, we assume that 
$\bm{v} = (v_1, v_2)$ is a complex and circularly-symmetric Gaussian process. 
Then

\begin{equation}
\label{eq:r_12_complex}
\begin{split}
    \hat{r}_{12}
    & = \langle (\hat{v}_{1r}+j\hat{v}_{1i})
        (\hat{v}_{2r}-j\hat{v}_{2i}) \rangle \\
    &=  \langle \hat{v}_{1r} \hat{v}_{2r} \rangle + 
        \langle \hat{v}_{1i} \hat{v}_{2i} \rangle + 
        j \left ( -\langle \hat{v}_{1r} \hat{v}_{2i} \rangle +  
        \langle \hat{v}_{1i} \hat{v}_{2r} \rangle \right )
\end{split}
\end{equation}

The circular symmetry of $\bm{v}$ implies that 
$\langle \hat{v}_{1r} \hat{v}_{2r} \rangle = 
\langle \hat{v}_{1i} \hat{v}_{2i} \rangle$ and
$-\langle \hat{v}_{1r} \hat{v}_{2i} \rangle = 
\langle \hat{v}_{1i} \hat{v}_{2r} \rangle$ so

\begin{equation}
\label{eq:r_12_complex_1}
\begin{split}
    \hat{r}_{12}
    &= 2\left[ \langle \hat{v}_{1r} \hat{v}_{2r} \rangle + 
        j\langle \hat{v}_{1i} \hat{v}_{2r} \rangle \right] \\
    &= 2(\hat{r}_{1r,2r}+j\hat{r}_{1i,2r})
\end{split}
\end{equation}

Now, for $\hat{r}_{1r,2r}=\langle \hat{v}_{1r} \hat{v}_{2r} \rangle$ and
$\hat{r}_{1i,2r}=\langle \hat{v}_{1i} \hat{v}_{2r} \rangle $ which are real,
we can use Equation \ref{eq:r_12_a} so

\begin{equation}
\label{eq:r_12_complex_2}
\begin{split}
    \hat{r}_{12}
    &= 2\left \{
       \left[
       r_{1r,2r} \left(1 + \frac{\langle v_{1r}e_{1r} \rangle }{\sigma_{1r}^2} + 
                     \frac{\langle v_{2r}e_{2r} \rangle }{\sigma_{2r}^2} \right)
       + \langle e_{1r}e_{2r} \rangle 
       \right] + j
       \left[
       r_{1i,2r} \left(1 + \frac{\langle v_{1i}e_{1i} \rangle }{\sigma_{1i}^2} + 
                     \frac{\langle v_{2r}e_{2r} \rangle }{\sigma_{2r}^2} \right)
       + \langle e_{1i}e_{2r} \rangle 
       \right]
       \right \} \\
    &= 2 \left( r_{1r,2r} + j r_{1i,2r}\right)
    \left(1 + \frac{\langle v_{1r}e_{1r} \rangle }{\sigma_{1r}^2} + 
              \frac{\langle v_{2r}e_{2r} \rangle }{\sigma_{2r}^2}\right)
        + 2 \left(\langle e_{1r}e_{2r} \rangle +
            j\langle e_{1i}e_{2r} \rangle \right)\\
    &= r_{12}
    \left(1 + \frac{\langle v_{1r}e_{1r} \rangle }{\sigma_{1r}^2} + 
              \frac{\langle v_{2r}e_{2r} \rangle }{\sigma_{2r}^2}\right)
        + 2 \left(\langle e_{1r}e_{2r} \rangle +
            j\langle e_{1i}e_{2r} \rangle \right)
\end{split}
\end{equation}

\noindent where in the second step we used the fact that 
$\langle v_{1r}e_{1r} \rangle/\sigma_{1r}^2=
\langle v_{1i}e_{1i} \rangle/\sigma_{1i}^2$ and in the third step
we used $r_{12}=2 \left( r_{1r,2r} + j r_{1i,2r}\right)$. All these follow from
circular symmetry. Note that all the terms in Equation \ref{eq:r_12_complex_2}
can be obtained from Equations \ref{eq:r_ve_1} and \ref{eq:r_e1_e2} using
$\sigma_{1r}^2=\sigma_{1i}^2=\sigma_1^2/2$ and 
$\sigma_{2r}^2=\sigma_{2i}^2=\sigma_2^2/2$.

We can use Equation \ref{eq:r_12_complex_2} to draw some important conclusions
regarding how quantization affects the magnitude and phase of $r_{12}$. 
We can write 

\begin{equation}
\label{eq:r_12_complex_3}
    \hat{r}_{12}
    = \alpha r_{12}+\beta, \hspace{0.5in} 
    \alpha = \left(1 + \frac{\langle v_{1r}e_{1r} \rangle }{\sigma_{1r}^2} + 
    \frac{\langle v_{2r}e_{2r} \rangle }{\sigma_{2r}^2}\right), \hspace{0.5in} 
    \beta = 2 \left(\langle e_{1r}e_{2r} \rangle +
            j\langle e_{1i}e_{2r} \rangle \right)
\end{equation}

Note that $\alpha$ is real, independent of $\rho$, and only contributes to the 
biasing of the magnitude of $\hat{r}_{12}$. On the other hand, $\beta$ is 
complex in general and affects both the magnitude and phase of $\hat{r}_{12}$.

Quantization will bias the magnitude of $\hat{r}_{12}$ except when 
$\alpha=1$ and $\beta=0$. This occurs approximately
when both inputs are optimally quantized since in this case
$\langle v_{1r}e_{1r} \rangle \rightarrow 0$, 
$\langle v_{2r}e_{2r} \rangle \rightarrow 0$ (so $\alpha \rightarrow 1$, see 
Section \ref{sec:real_quantizer} and Figure 
\ref{fig:real_quantization_15_16_levels}), 
and also $\langle e_{1r}e_{2r} \rangle \rightarrow 0$, 
$\langle e_{1i}e_{2r} \rangle \rightarrow 0$ (so $\beta \rightarrow 0$, see
Section \ref{sec:real_correlator} and Figure 
\ref{fig:real_corr_e1e2_15_16_levels}).

Quantization will bias the phase of $\hat{r}_{12}$ except in two cases:
the first case is when $\beta=0$, which occurs approximately 
when at least one of the  
inputs is optimally quantized (see Section \ref{sec:real_correlator} and Figure
\ref{fig:real_corr_e1e2_15_16_levels}). Note that this is a less stringent 
requirement than that for unbiased magnitude, which requires both inputs to be
optimally quantized.

The second case for negligible bias in the phase of $\hat{r}_{12}$ 
occurs when $\rho \ll 1$ since using Equation 
\ref{eq:r_e1_e2_small_rho} in Equation \ref{eq:r_12_complex_2} we have

\begin{equation}
\label{eq:r_12_complex_small_rho}
\begin{split}
    \left. \hat{r}_{12} \right|_{\rho \ll 1}  
    &\approx 2r_{1r,2r} 
        \left(1 + \frac{\langle v_{1r}e_{1r} \rangle }{\sigma_{1r}^2} \right) 
        \left(1 + \frac{\langle v_{2r}e_{2r} \rangle }{\sigma_{2r}^2}\right)
        + 2jr_{1i,2r} 
        \left(1 + \frac{\langle v_{1i}e_{1i} \rangle }{\sigma_{1i}^2} \right) 
        \left(1 + \frac{\langle v_{2r}e_{2r} \rangle }{\sigma_{2r}^2} \right)\\
    &= r_{12}
        \left(1 + \frac{\langle v_{1r}e_{1r} \rangle }{\sigma_{1r}^2} \right) 
        \left(1 + \frac{\langle v_{2r}e_{2r} \rangle }{\sigma_{2r}^2} \right).
\end{split}
\end{equation}

Since the factors that multiply $r_{12}$ are real then 
$\angle \left(\hat{r}_{12} \right)=\angle \left(r_{12} \right)$.

\begin{figure}[htbp]
    \centering
    \includegraphics[width=\textwidth]{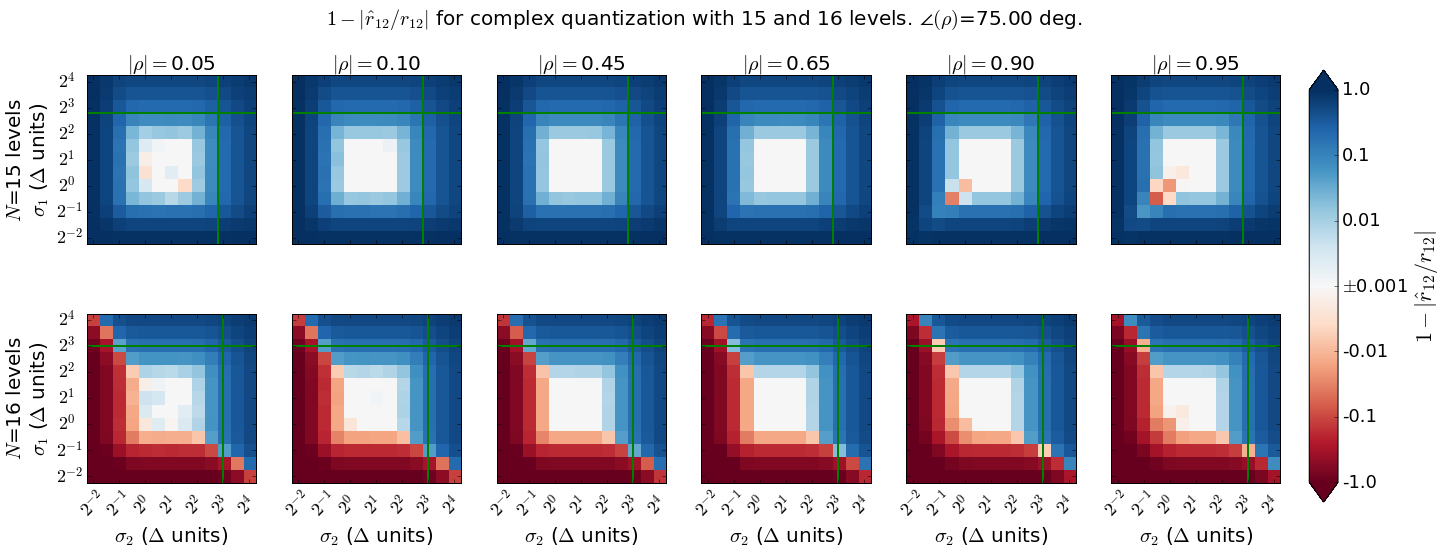}
    \caption{
$|\hat{r}_{12}/r_{12}|$ from simulations as function of 
$\sigma_1$ and $\sigma_2$ for different values of $|\rho|$.
For moderate values of $\rho$
the bias in the magnitude of $|\hat{r}_{12}|$ is below $\sim 0.1\%$ 
within the inner square enclosed by the region 
$\sigma_1 \times \sigma_2 \approx [2^{-0.1}, 2^{1.4}] \times 
[2^{-0.1}, 2^{1.4}]$, 
corresponding to region where both inputs are optimally quantized.
For $|\rho|\gtrsim 0.85$ this bias can increase up to $\sim 4\%$.
    }
    \label{fig:complex_corr_mag_15_16_levels}
\end{figure}

\begin{figure}[htbp]
    \centering
    \includegraphics[width=\textwidth]{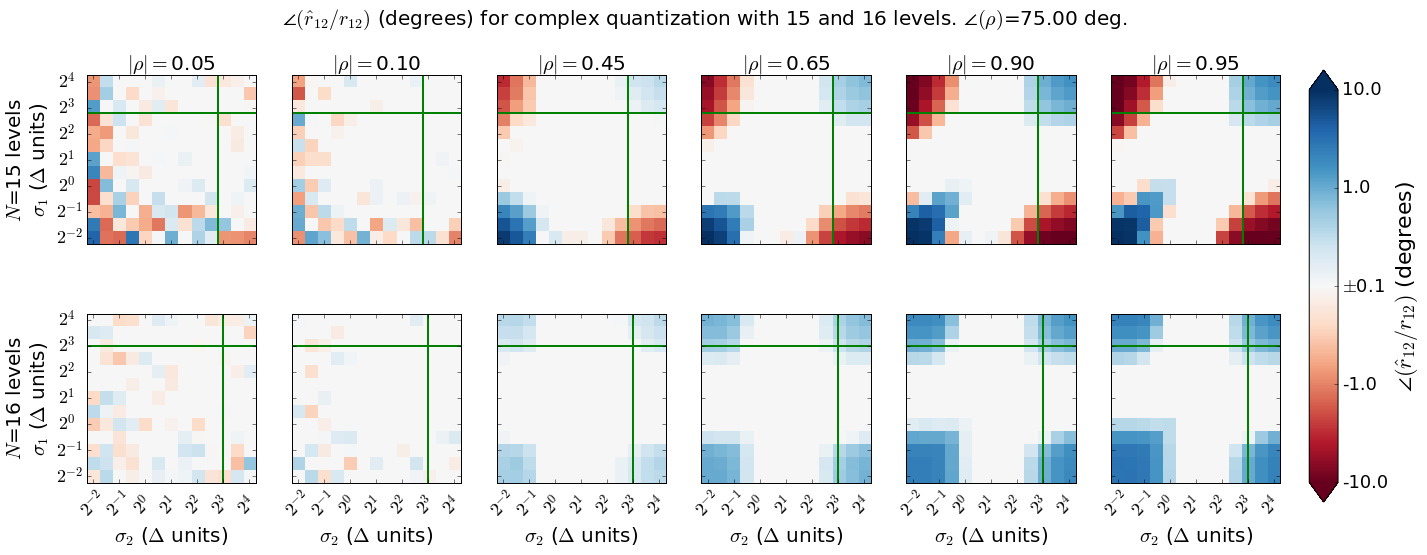}
    \caption{
$\angle (\hat{r}_{12}/r_{12})$ (in degrees) from simulations as 
function of $\sigma_1$ and $\sigma_2$ for different values of $|\rho|$.
The bias in the phase of $\hat{r}_{12}$ is negligible when at least one of the 
inputs is optimally quantized. This is a less stringent 
requirement than that for the magnitude, which requires both inputs to be
optimally quantized. The bias is
below $\sim 0.1^\circ$ within the cross-shaped region where either 
$\sigma_1$ or $\sigma_2$ are in the approximate interval 
$[2^{-0.1}, 2^{1.4}]$. When $|\rho|$ is high (last two columns)
the bias within this region can rise up to $\sim 1^\circ$.
When $|\rho|$ is small
(first two columns) 
the phase bias is below $\sim 0.1^\circ$ for all values of $\sigma_1$ and 
$\sigma_2$, although there are still random fluctuations in the simulation at 
the $\sim$ sub-degree level for very low values of $\sigma$ (see text).
    }
    \label{fig:complex_corr_phase_15_16_levels}
\end{figure}

Figures \ref{fig:complex_corr_mag_15_16_levels} and
\ref{fig:complex_corr_phase_15_16_levels} show results from simulations
of $\hat{r}_{12}/r_{12}$ 
(magnitude and phase respectively. The phase is in degrees). The method is the 
same as in Section \ref{sec:real_quantizer}, but this time the $10^7$ sample 
vectors $(v_1[n], v_2[n])$ are drawn from a circularly symmetric Gaussian 
distribution. We only vary the magnitude of $\rho$, keeping its phase fixed at 
75 degrees.

Note that Equations \ref{eq:r_12_complex_2}-\ref{eq:r_12_complex_small_rho}
predict accurately the behavior of the magnitude and phase of $\hat{r}_{12}$.
For moderate values of $\rho$ ($|\rho|\lesssim 0.85$)
the bias in the magnitude (Figure 
\ref{fig:complex_corr_mag_15_16_levels}) is below $\sim 0.1\%$ roughly within
the inner square enclosed by the region 
$\sigma_1 \times \sigma_2 \approx [2^{-0.1}, 2^{1.4}] \times 
[2^{-0.1}, 2^{1.4}]$, 
corresponding to region where both inputs are optimally quantized (see
Section \ref{sec:complex_quantizer}). 
For $|\rho|\gtrsim 0.85$ the bias within this region can increase up to $\sim 4\%$.

As for the phase
(Figure \ref{fig:complex_corr_phase_15_16_levels}), the bias is
below $\sim 0.1^\circ$ within the cross-shaped region where either 
$\sigma_1$ or $\sigma_2$ are optimally quantized.
When $|\rho|\gtrsim 0.85$ the bias within this region can rise up to $\sim 1^\circ$.
When $\rho \lesssim 0.1$
(first two columns of Figure \ref{fig:complex_corr_phase_15_16_levels}) 
the phase bias is below $\sim 0.1^\circ$ (values from Equation~\ref{eq:r_12_complex_2})
for all values of $\sigma_1$ and $\sigma_2$ as
predicted by Equation \ref{eq:r_12_complex_small_rho}, although there are 
still random fluctuations in the simulation at the $\sim$ sub-degree level for 
very low values of $\sigma$ (bottom and left edges of the plots) for reasons 
explained in Section \ref{sec:real_correlator}.

%###############################################################################
\section{Implications for radio interferometry}
\label{sec:interferometry}

The results above have important implications for radio interferometry.
Quantization will have a significant biasing effect on the visibility 
magnitude unless both inputs are optimally quantized, which can be a stringent
requirement (both signal levels need to be in the region where the uncorrelated 
quantization model is valid). However, we have found that the
bias in the visibility phase is negligible
even in conditions as extreme as when one of the inputs is 
suffering from severe clipping, or even when both inputs are severely clipped in 
the case of weak sources ($|\rho| \ll 1$). The same conditions apply when one or 
both input levels are very low (note that any of these extreme conditions
will affect the signal-to-noise ratio of the measured 
visibility even if the phase is unbiased, but that analysis is beyond the scope 
of this paper). An accurate determination of the visibility phase is 
critical for beamforming, fringe stopping, and image reconstruction techniques.

For the particular case of CHIME, in which the sky signals are weak
and the correlator inputs are dominated by the noise of the analog receiving 
system, the correlation coefficient is typically low ($|\rho| \lesssim 0.1$)
even for the brightest radio point sources such as CasA, CygA, and TauA, but 
excluding the sun.
This means that, except for the time when the sun is in the 
primary beam of the CHIME telescope ($\sim 14$ minutes per day), all the 
visibility phases will have negligible bias due to quantization. 

The quantization 
bias also has an effect on the beamformed sensitivity of a radio 
interferometric array. To illustrate this, consider a one-dimensional array 
consisting of uniformly spaced feeds located at positions
$0, 1, \ldots, N_{f}-1$, in units of the normalized feed spacing
$b_{\lambda} = b/\lambda$, where $\lambda$ is the observed wavelength. 
This example corresponds to one of the cylinders of 
the CHIME telescope, where the feeds are uniformly spaced along the axis of the
cylinder. The cylinder axis (and thus the linear array) is oriented North-South 
(N-S), so the resolution in the N-S 
direction is provided by the correlations between feeds.
We will assume that all the feeds have identical beams that are N-S isotropic 
and receivers with system noise
$\sigma_{sys}^2$, although the generalization is straightforward.

For a point source on the meridian with noise temperature $\sigma^2$ such that 
the signal-to-system-noise ratio is $SNR=\sigma^2/\sigma_{sys}^2$, the 
unquantized autocorrelations for each feed are identical and equal to 

\begin{equation}
\label{eq:autos}
    r_{ii} = \sigma_{sys}^2 (1+SNR), \hspace{0.5in} i=0, 1, \ldots, N_{f}-1
\end{equation}

\noindent while the unquantized visibility and correlation coefficient between 
feeds $i$ and $j$ are

\begin{equation}
\label{eq:visibility}
    r_{ij} = SNR \cdot \sigma_{sys}^2  e^{-j 2 \pi (i-j) b_{\lambda} \sin \theta}, 
    \hspace{0.5in} 
    \rho_{ij} = \frac{e^{-j2\pi (i-j)b_{\lambda} \sin \theta}}{1+\frac{1}{SNR}}
\end{equation}

\noindent where $\theta$ is the source zenith angle  
and we have assumed uncorrelated system noise between
feeds.

\begin{figure}[htbp]
    \centering
    \includegraphics[width=\textwidth]{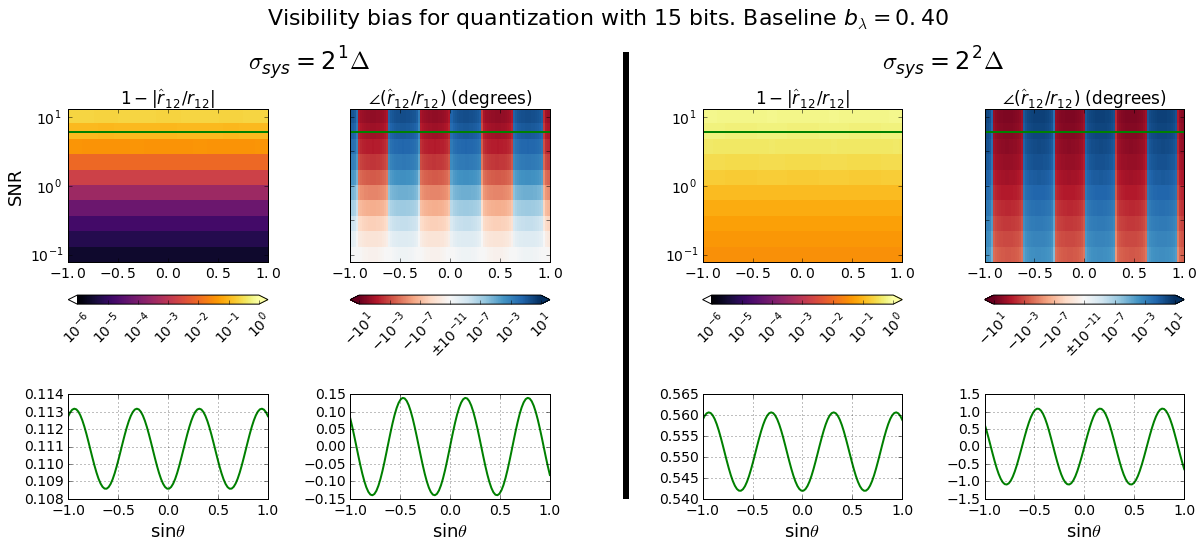}
    \caption{
Bias due to quantization of the measured visibility as function of
the source position $\theta$ and the signal-to-system-noise ratio is $SNR$.
The visibility baseline is $b_\lambda=0.4$ which corresponds
to the CHIME normalized feed spacing at 400 MHz. When 
$\sigma_{sys} = 2^{1} \Delta$ (left panels), well within the optimal 
quantization interval, the quantization bias for weak sources 
($SNR\lesssim 0.1$) is negligible. This is the 
regime for CHIME $\sim 99\%$ of the time. When 
$\sigma_{sys} = 2^{1} \Delta$ (right panels), 
which is the optimal input level according to the 
uncorrelated quantization noise model, the amount of bias increases significantly.
    }
    \label{fig:visibility_bias}
\end{figure}

The bias due to quantization of the measured visibility as function of 
$SNR$ and $\theta$ for $i=j+1$ (consecutive feeds) is shown in 
Figure \ref{fig:visibility_bias}. We use $b_\lambda=0.4$ which corresponds
to the CHIME normalized feed spacing at 400 MHz. These results are obtained 
directly from Equation \ref{eq:r_12_complex_2}.

To illustrate the difference between Equation \ref{eq:r_12_complex_2} and 
the uncorrelated quantization noise model, and the importance of optimizing the 
input signal level of the quantizer, Figure \ref{fig:visibility_bias} shows
the bias for two different values of $\sigma_{sys}$: $2^1 \Delta$ (left panels) and
$2^2 \Delta$ (right panels). For a system-noise dominated telescope like CHIME, 
the correlator inputs are calibrated so $\sigma_{sys}$ corresponds to the
optimal input level of the quantizer in order to minimize the effects of 
quantization. For $N=15$, the optimal input level according to the uncorrelated 
quantization noise model is $\sigma_{sys} \approx 2^2 \Delta$, corresponding to the 
point where $\sigma_e$ is minimum (see second row of Figure 
\ref{fig:complex_quantization_15_16_levels}). On the other hand, Equation 
\ref{eq:r_12_complex_2} suggests that a better choice for $\sigma_{sys}$ should
be more centered around the optimal quantization interval 
$[2^{-0.1}\Delta, 2^{1.4}\Delta]$. The CHIME digital calibration 
module uses $\sigma_{sys}\approx 2^{1} \Delta$, which is well within
this interval while still keeping $\sigma_e$ relatively low 
(see second and third rows of Figure 
\ref{fig:complex_quantization_15_16_levels}, if $\sigma_{sys}$ is too close
to the lower end of the interval then the contribution of $\sigma_e$ is
significant).

Note that the $SNR$ sets the overall 
amount of bias due to quantization since this parameter defines both 
$r_{ii}$ and $|\rho_{ij}|$ (Equations \ref{eq:autos} and \ref{eq:visibility}). 
For $\sigma_{sys} = 2^{1} \Delta$
and $|\rho_{ij}| \lesssim 0.1$ (so $SNR \lesssim 0.1$) the magnitude bias is
$\lesssim 10^{-6}$ and the phase bias is $\lesssim 10^{-11}$ degrees,
too small to have any significant impact that requires the
generalized Van Vleck correction from \citet{2016arXiv160804367B}. 
As mentioned before, this is the 
regime for CHIME $\sim 99\%$ of the time. However, when the sun is in the main 
beam ($\sim 1\%$ of the time), the $SNR$ can be as high as $\sim 6$ (green
line in Figure \ref{fig:visibility_bias}), 
corresponding to a magnitude bias of $\sim 11\%$ and a phase bias of
up to $\sim 0.15^\circ$. Although the CHIME cosmology
data pipeline masks out the sun time, this data is still very useful for beam 
mapping purposes. 
The quantization bias is significant enough in this case to justify the
implementation of the generalized Van Vleck correction\footnote{Note that we 
are assuming that the sun is a point
source to simplify the analysis since we are interested in 
studying the behavior of quantization for strong sources. Although, strictly
speaking, the sun is an extended source for CHIME, for observations with the 
CHIME pathfinder (a small version of CHIME with 256 receivers and 10\% of the
full instrument collecting area) this is an adequate approximation.}. 

When $\sigma_{sys} = 2^{2} \Delta$ (right side of Figure 
\ref{fig:visibility_bias}) the amount of bias increases significantly even
in the weak-source regime. For $SNR \sim 0.1$ the magnitude bias is
$\sim3\%$ and the phase bias is $\sim 3 \times 10^{-3}$ degrees, while for $SNR \sim 6$
the magnitude bias is $\sim56\%$ and the phase bias is $\sim 1^\circ$, demonstrating
that for this particular application the uncorrelated quantization noise 
model must be used carefully since it can introduce important effects
in the measured visibilities.

The quantization bias also depends on the position of the source and the
baseline. These parameters determine $\angle(\rho_{ij})$ which affects
the measured visibility $\hat{r}_{ij}$ through the second term of Equation
\ref{eq:r_12_complex_2}. As Figure~\ref{fig:visibility_bias} shows, the 
position dependence manifests as fringes as a function of $l=\sin \theta$,
where the baseline determines the quantization fringe rate.

We can use the $N_{f}(N_{f}-1)/2$ visibilities (excluding the 
autocorrelations) to beamform in the direction of the source. 
Since for $k=i-j$ fixed there are
$(N_{f}-k)$ identical baselines, then we can write the quantized beamformed 
output as

\begin{equation}
\label{eq:beamformed_q}
\begin{split}
    \hat{R} &= \sum_{i>j}^{N_{f}-1} \hat{r}_{ij} e^{j 2 \pi (i-j) b_{\lambda} \sin \theta}
      = \sum_{k=1}^{N_{f}-1} (N_{f}-k)
        \hat{r}_{k} e^{j 2 \pi k b_{\lambda} \sin \theta}
\end{split}
\end{equation}

\noindent while the unquantized beamformed output is

\begin{equation}
\label{eq:beamformed}
\begin{split}
    R &= \sum_{k=1}^{N_{f}-1} (N_{f}-k)r_{k} e^{j 2 \pi k b_{\lambda} \sin \theta}
       = \frac{N_f(N_f-1)}{2}SNR \sigma_{sys}^2.
\end{split}
\end{equation}

\begin{figure}[htbp]
    \centering
    \includegraphics[width=\textwidth]{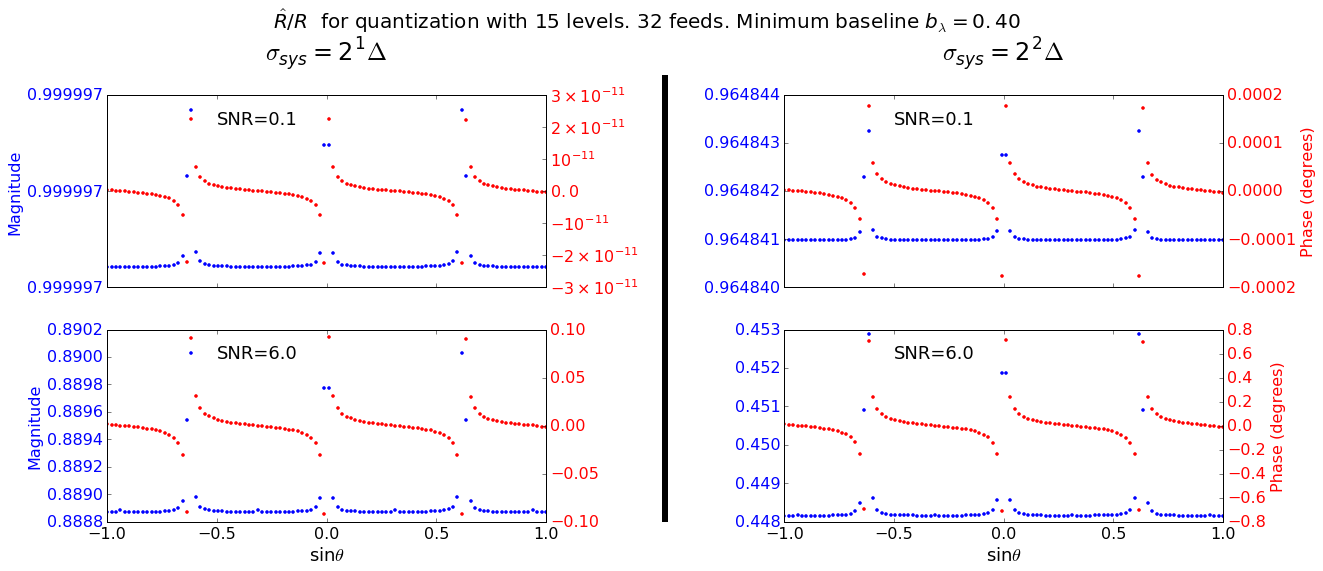}
    \caption{
Complex quantization parameter $\eta_q =\hat{R}/R$
as function of the source position $\theta$
for $SNR= 0.1$ (top row, this $SNR$ is the approximate upper limit of 
weak-source regime) and $SNR= 6$ 
(bottom row, this is the typical $SNR$ of a strong source like the sun). 
For each plot, the blue labels and dots correspond to the magnitude of $\eta_q$
and the red labels and dots correspond to its phase in degrees.
Note that for $\sigma_{sys}=2^1 \Delta$ 
(left column), which is well within the optimal quantization interval for 
$N=15$ levels, and in the weak-source regime ($SNR\lesssim 0.1$, top left plot),
$\eta_q$ is very close to being real-valued and
deviates from unity by less than one part in $\sim 3\times 10^{-6}$ so the loss 
of beamforming efficiency due to quantization is negligible. 
If we set $\sigma_{sys}=2^2 \Delta$ (right column), the beamforming sensitivity
reduces significantly even in the weak-source regime. This confirms that for
this application the uncorrelated quantization model leads to important 
deviations from the expected performance of the interferometric array.
    }
    \label{fig:beamforming_efficiency}
\end{figure}

We can define a complex quantization parameter 

\begin{equation}
\label{eq:beamforming_efficiency}
    \eta_q = \frac{\hat{R}}{R}
\end{equation}

\noindent as a measure of the beamforming efficiency due to quantization. 
Figure \ref{fig:beamforming_efficiency} shows the
magnitude and phase of $\eta_q$ as function of the source position $\theta$
for $SNR= 0.1$ (approximate upper limit of weak-source regime) and $SNR= 6$ 
(typical strong source like the sun). We used $N_f=32$ and kept 
$b_\lambda$ fixed at 0.4.

The most important feature from Figure \ref{fig:beamforming_efficiency} is 
that for $\sigma_{sys}=2^1 \Delta$ (left column) and in the weak-source regime 
($SNR\lesssim 0.1$, top left plot) the loss of beamforming efficiency due to 
quantization is negligible ($\eta_q$ is very close to being real-valued and
deviates from unity by less than one part in $\sim 3\times 10^{-6}$).
However, for a strong source like the sun the beamforming
efficiency decreases below $\sim 89 \%$ (bottom left plot). When $\sigma_{sys}$ 
is set to $2^2 \Delta$ (right column)
the beamforming sensitivity reduces to $\sim 96 \%$ and $\sim 45 \%$
for $SNR = 0.1$ and $6$ respectively, confirming that the 
uncorrelated quantization noise model leads to important deviations from
the expected interferometer performance.

%###############################################################################
\section{Conclusions}
\label{sec:conclusions}

We investigated the correlation between the input and the quantization 
error of a quantizer with uniformly spaced levels and an odd symmetric transfer 
function. We then used these results to explore the biasing effect of 
quantization in the correlation measured by a complex-valued digital correlator.

We showed that, for a complex-valued quantizer with a circularly symmetric
Gaussian input, the correlation between the input and the quantization error is 
always real. It is always negative when the number of levels $N$ of the 
quantizer is odd, while for $N$ even this correlation is 
positive in the low signal level regime. 
In both cases there is an interval for the signal level $\sigma$ (which we 
denote the interval of optimal quantization) for which this input-error 
correlation is very weak and the uncorrelated quantization error model provides 
a very accurate approximation. The length of the optimal quantization 
interval depends on $N$ and on the tolerance required by each specific application.

With these results we determined the quantization bias in the correlations 
measured by a digital correlator and derived the conditions under which the
bias in the magnitude and phase of the measured correlation is
negligible with respect to the unquantized values: we demonstrated that the 
magnitude bias is negligible only if both unquantized inputs are 
optimally quantized, 
while the phase bias is negligible when 1) at least one of the  
inputs is optimally quantized, or when 2) the correlation coefficient $\rho$
between the unquantized inputs is small.

These results are important for radio interferometry where the 
correlations measured by the digital correlator provide the 
interferometric visibilities. Although quantization will bias 
significantly the visibility
magnitude unless both inputs are optimally quantized, which can be a stringent
requirement, we showed that the bias in the visibility phase is negligible
even in extreme conditions like when one of the inputs is in the high-$\sigma$
regime with large amounts of clipping or when it is in the low-$\sigma$ regime
where the contribution of the quantization error to the quantized output 
is very high. Even when both inputs are far from the optimal quantization
regime (either because of extreme clipping or very low signal level) the 
phase quantization bias is negligible for weak sources ($|\rho| \ll 1$). 
This is the typical case for interferometers like CHIME where the analog
inputs are dominated by the receiver noise. In this regime all the visibility
phases will be approximately unbiased regardless of the signal levels.

Finally, we demonstrated using a specific example corresponding to a
CHIME-like array of antennas that quantization 
reduces the point-source sensitivity of a radio interferometric array.
For a system-noise dominated telescope like CHIME, this effect can be reduced
to negligible levels in the weak-source regime with a suitable scaling of 
the system noise level at the input of the quantizer.

Highly redundant telescopes like CHIME are becoming more common in present and
future observatories. The detailed analysis and knowledge of this paper will
serve to optimize the calibration and digitization of these instruments.

%###############################################################################
\section{Acknowledgements}
\label{sec:acknowledgements}

We thank James Moran, Bernard Widrow, and the members of the CHIME 
collaboration for their comments and stimulating discussions. We acknowledge 
funding from the Natural Sciences and 
Engineering Research Council of Canada, Canadian Institute for Advanced 
Research, Canadian Foundation for Innovation, and le 
Cofinancement gouvernement du Qu\'{e}bec-FCI.

%###############################################################################
%\appendix
\appendix{$\langle ve \rangle$ for a real quantizer}
\label{app:r_ve}

Here we derive Equation \ref{eq:r_ve_1}

\begin{equation}
\label{eq:r_ve_1_app1}
\begin{split}
    \langle ve \rangle 
    & = \sum_{i=1}^N \int_{y_{i-1}}^{y_i} (k_i-v)v \mathcal{N}(v|\sigma^2) dv .
\end{split}
\end{equation}

Evaluating the integral and re-arranging 

\begin{equation}
\label{eq:r_ve_1_app2}
\begin{split}
    \langle ve \rangle 
    & = \sigma^2 \sum_{i=1}^N \left \{
        \left(y_i-k_i\right) \mathcal{N}(y_i|\sigma^2) -
        \left(y_{i-1}-k_i\right) \mathcal{N}(y_{i-1}|\sigma^2) -
        \frac{1}{2}\left[ \text{erf}\left( \frac{y_i}{\sqrt{2\sigma^2}} \right)- 
                        \text{erf}\left( \frac{y_{i-1}}{\sqrt{2\sigma^2}}\right)
                    \right] \right \} .
\end{split}
\end{equation}

The summation of the erf terms in the square brackets gives 2. As for the first 
two terms of Equation \ref{eq:r_ve_1_app2}, note that $y_{i}-k_i=1/2$ and 
$y_{i-1}-k_i=-1/2$. Simplifying we obtain

\begin{equation}
\label{eq:r_ve_1_app3}
\begin{split}
    \langle ve \rangle 
    & = \sigma^2 \left[-1 + \sum_{i=1}^{N-1} 
            \mathcal{N}\left(\left.-\frac{N}{2}+i\right|\sigma^2\right)\right] .
\end{split}
\end{equation}

Since $\mathcal{N}(v|\sigma^2)$ is an even function we can also write

\begin{equation}
\label{eq:r_ve_1_app4}
    \langle ve \rangle = \begin{cases} \displaystyle
        \sigma^2 \left[-1 + 2\sum_{i=0}^{\frac{N-3}{2}} 
            \mathcal{N}\left(\left.\frac{1}{2}+i\right|\sigma^2\right) \right]
            & \text{if $N$ odd}\\ \displaystyle
        \sigma^2 \left[-1 + \frac{1}{\sqrt{2\pi\sigma^2}} + 2\sum_{i=0}^{\frac{N-4}{2}} 
            \mathcal{N}\left(\left.1+i\right|\sigma^2\right) \right]
            & \text{if $N$ even}  
        \end{cases}          
\end{equation}

\noindent where it is clear that the summation term is zero for $N=2$. 

To find $\sigma_e^2 = \langle e^2 \rangle$ in Equation \ref{eq:var_e} we
follow the same procedure

\begin{equation}
\label{eq:var_e_app}
\begin{split}
    \sigma_e^2
    & = \sum_{i=1}^N \int_{y_{i-1}}^{y_i} (k_i-v)^2 \mathcal{N}(v|\sigma^2)dv
      = -\langle ve \rangle + 
       \sum_{i=1}^N \int_{y_{i-1}}^{y_i} k_i(k_i-v) \mathcal{N}(v|\sigma^2)dv.
\end{split}
\end{equation}

Evaluating the integral

\begin{equation}
\label{eq:var_e_app1}
\begin{split}
    \sigma_e^2
    & = -\langle ve \rangle + \sum_{i=1}^N 
        \left \{
        \frac{k_i^2}{2}\left[\text{erf}\left(\frac{y_i}{\sqrt{2\sigma^2}}\right) 
                        -\text{erf}\left(\frac{y_{i-1}}{\sqrt{2\sigma^2}}\right)
                       \right] 
        + k_i \sigma^2  \left[ \mathcal{N}(y_i|\sigma^2) - 
                              \mathcal{N}(y_{i-1}|\sigma^2)
                        \right]            
        \right \}.
\end{split}
\end{equation}

The summation of the second term inside the curly brackets gives
$-\sigma^2\sum_{i=1}^{N-1}\mathcal{N}(y_i|\sigma^2) = 
-\langle ve\rangle-\sigma^2$. As for the erf terms, after re-arranging
we obtain

\begin{equation}
\label{eq:var_e_app2}
\begin{split}
    \sigma_e^2
    & = -2\langle ve\rangle -\sigma^2 + \left(\frac{N-1}{2}\right)^2
        - \sum_{i=1}^{N-1} \frac{1}{2} \left(k_{i+1}^2-k_{i}^2\right)          
        \text{erf}\left( \frac{y_{i}}{\sqrt{2\sigma^2}} \right).
\end{split}
\end{equation}

Finally, using Equation \ref{eq:q_levels} we obtain

\begin{equation}
\label{eq:var_e_app3}
\begin{split}
    \sigma_e^2
    & = -2\langle ve\rangle -\sigma^2 + \left(\frac{N-1}{2}\right)^2
        - \sum_{i=1}^{N-1} \left(-\frac{N}{2}+i\right)          
       \text{erf}\left(\frac{-N/2+i}{\sqrt{2\sigma^2}}\right).
\end{split}
\end{equation}

Equation \ref{eq:var_vo} in Section \ref{sec:real_quantizer} follows from the 
two results above.

%###############################################################################
\appendix{Sign of $\langle ve \rangle$}
\label{app:sign_r_ve}
 
To show that $\langle ve \rangle$ in Equation \ref{eq:r_ve_1} is always 
negative for a quantizer with an odd number of levels ($N$ odd), it is enough 
to show that 
$S_o = \sum_{i=0}^{M} \mathcal{N}\left(\left. 1/2+i\right|\sigma^2\right)
<1/2$ for all $M$ positive integer and $\sigma>0$ real (it is clear that
$S_o>0$). Note that

\begin{equation}
\begin{split}
\label{eq:So_1}
S_o &< \sum_{i=0}^\infty \frac{1}{\sqrt{2\pi\sigma^2}}e^{-(1/2+i)^2/(2\sigma^2)}
    = \frac{\vartheta_2(q)}{2\sqrt{2\pi\sigma^2}}
\end{split}
\end{equation}

\noindent where $\vartheta_2$ is the Jacobi theta function\footnote{For details see 
\href{http://mathworld.wolfram.com/JacobiThetaFunctions.html}
{\texttt{http://mathworld.wolfram.com/JacobiThetaFunctions.html}}} and 

\begin{equation}
\begin{split}
\label{eq:q}
q = e^{-1/(2\sigma^2)}=e^{-\pi K^\prime/K}
\end{split}
\end{equation}

\noindent where $K(k)$ is the complete elliptic integral of the first kind\footnote{For 
details see 
\href{http://mathworld.wolfram.com/CompleteEllipticIntegraloftheFirstKind.html}
{\texttt{http://mathworld.wolfram.com/CompleteEllipticIntegraloftheFirstKind.html}}}, 
$k$ is the elliptic modulus, and $K^\prime(k)=K(\sqrt{1-k^2})$. 
The functions $\vartheta_2(q)$ and $K(k)$ are related through 
$\vartheta_2^2(q) = 2kK(k)/\pi$. Also, from Equation \ref{eq:q} we have 
$\sigma^2 = K/(2\pi K^\prime)$. Using these results in Equation \ref{eq:So_1}, 
we find that

\begin{equation}
\begin{split}
\label{eq:So_2}
S_o < \frac{1}{2} \sqrt{\frac{2kK^\prime}{\pi}}.
\end{split}
\end{equation}

The function $f(k)=2kK^\prime/\pi$ is a strictly increasing function of $k$ and 
maps the $k$-interval $(0, 1)$ (corresponding to $\sigma \in (0, \infty)$) to 
the interval $(0, 1)$. Thus $f(k)<1$ in this interval and it 
follows that $S_o<1/2$.

For the quantizer with an even number of levels ($N$ even), note that the sum
$S_e = \sum_{i=0}^{M} \mathcal{N}\left(\left. 1+i\right|\sigma^2\right)$ is 
just the right Riemann sum of
$\mathcal{N}\left(\left. v\right|\sigma^2\right)$ over the interval $[0, M]$. 
Since $\mathcal{N}$ is a strictly decreasing function over this interval then
it follows that 

\begin{equation}
\begin{split}
\label{eq:Se_1}
S_e < \int_0^{M} \frac{1}{\sqrt{2\pi\sigma^2}}e^{-v^2/(2\sigma^2)} dv
    = \frac{1}{2}\text{erf}\left(\frac{M}{\sqrt{2\sigma^2}}\right) <\frac{1}{2}.
\end{split}
\end{equation}

Thus, the summation (last) term of Equation \ref{eq:r_ve_1} for $N$ even is 
also positive and bounded above by 1/2. Since the term $1/(\sqrt{2\pi\sigma^2})$ 
of this Equation becomes arbitrarily large as $\sigma$ decreases, 
then $\langle ve \rangle$ eventually becomes positive for $N$ even in the 
low-$\sigma$ regime.

%-------------------------------------------------------------------------------
%%%%% References %%%%%
%%\section*{References}

%%% I use: http://scieng.library.ubc.ca/coden/
%%% for Journal Title Abbreviations

%\bibliographystyle{plain}
\bibliographystyle{ws-jai}
\bibliography{bibliographybib}   % bibliography data

\end{document}